\documentclass[twocolumn,showpacs,preprintnumbers,amsmath,amssymb]{revtex4}

\def\undersim#1{\setbox9\hbox{${#1}$}{#1}\kern-\wd9\lower
    2.5pt \hbox{\lower\dp9\hbox to \wd9{\hss $_\sim$\hss}}}
\usepackage{amssymb}
\usepackage{amsmath}
\usepackage{graphicx}
\usepackage{color}

\def\undersim#1{\setbox9\hbox{${#1}$}{#1}\kern-\wd9\lower
    2.5pt \hbox{\lower\dp9\hbox to \wd9{\hss $_\sim$\hss}}}

\def\mr{{\mathbf r}}

\def\mr{{\mathbf r}}

\def\mk{{\mathbf p}}

\begin{document}

\title{Influence of in-medium mass width on Hanbury Brown-Twiss correlation
strength}

\author{Peng-Zhi Xu$^1$}
\author{Wei-Ning Zhang$^{1,2}$\footnote{wnzhang@hit.edu.cn, wnzhang@dlut.edu.cn}}
\affiliation{$^1$Department of Physics, Harbin Institute of Technology, Harbin,
Heilongjiang 150006, China\\
$^2$School of Physics, Dalian University of Technology, Dalian, Liaoning 116024,
China}

%\date{\today}

\begin{abstract}
The interactions of particles with medium may lead to a nonzero in-medium mass 
width of the particles and cause them to have different masses in the medium. 
In this letter, we investigate the influence of the in-medium mass width on 
the strength of Hanbury Brown-Twiss (HBT) correlation in high-energy heavy-ion 
collisions. It is found that the strength decreases with in-medium mass width 
of identical bosons. This influence are more significant for the boson with 
heavier mass and decreases with increasing particle momentum. 
\end{abstract}
\maketitle

Hanbury Brown-Twiss (HBT) correlation, or identical boson intensity correlation,
has been widely used in high-energy heavy-ion collisions to study the space-time
structure and coherence of the particle-emitting sources
\cite{Gyu79,Wongbook,Wie99,Wei00,Csorgo02,Lisa05}.
The strength of two-boson HBT correlation, i.e. the intercept of the correlation
term at zero relative momentum, should be one for a chaotic particle-emitting
source. However, except for coherent particle emission, there are many other
effects such as particle misidentification, final-state Coulomb repulsive force,
long-lived resonance decay, and so on can change the intercept, and a
$\lambda$-parameter is introduced to describe the intercept in data analyses
\cite{Gyu79,Wongbook,Wie99,Wei00,Csorgo02,Lisa05}.

In the particle-emitting sources produced in high-energy heavy-ion collisions,
the in-medium mass modification of bosons caused by the interactions between
the bosons and source medium may lead to a squeezed back-to-back correlation
of boson-antiboson pairs \cite{{AsaCso96,AsaCsoGyu99,Padula06,DudPad10,Zhang15,
Zhang15a,Zhang-EPJC16,AGY17,XuZhang19,XuZhang19a}}.
For identical bosons with nonzero in-medium mass width, they have different
masses in the source. In this point, their ``identicity" in the source is
reduced. Therefore, the finally observed HBT correlation strength may change
for the identical bosons. In this work, we examine the influence of the
in-medium mass width on the HBT-correlation strength with the technique of
quantum path integral of possibility amplitude \cite{Wongbook}. We find that
the strength decreases with the in-medium mass width and the influence is 
more significant for the boson with heavier mass and smaller momentum.

Tow-particle HBT correlation function is defined by
\begin{equation}
C(\mk_1,\mk_2)=\frac{P_2(\mk_1,\mk_2)}{P_1(\mk_1)P_1(\mk_2)},
\end{equation}
where $P_1$ and $P_2$ are observed single and double identical boson momentum
distributions.

For a boson produced at the source point $x$ with momentum $\mk$ and in-medium 
mass $m'$ and detected at $x_d$, the possibility amplitude can be expressed with
quantum path-integral method as \cite{Wongbook}
\begin{eqnarray}
&&\psi_1(\mk\!:x\buildrel m'\over\longrightarrow x_f\buildrel m\over
\longrightarrow x_d)={\cal A}(p',x)e^{i\phi(x)}~~~~~~~~~~~~~~~~~~\nonumber\\
&&\hspace*{38mm}\times\, e^{ip'\cdot(x-x_f)} e^{ip\cdot(x_f-x_d)},
\label{psi-1}
\end{eqnarray}
where $x_f$ is the freeze-out point, $m$ is the mass of the boson at free state,
${\cal A}$ and $\phi$ are the production amplitude and phase, $p'=(E'_\mk,\mk)$
and $p=(E_\mk,\mk)$ are the four-momenta of the quasiparticle and free particle, $E'_\mk=\sqrt{\mk^2+m'^2}$, and $E_\mk=\sqrt{\mk^2+m^2}$.

\begin{figure}[htbp]
\includegraphics[scale=0.57]{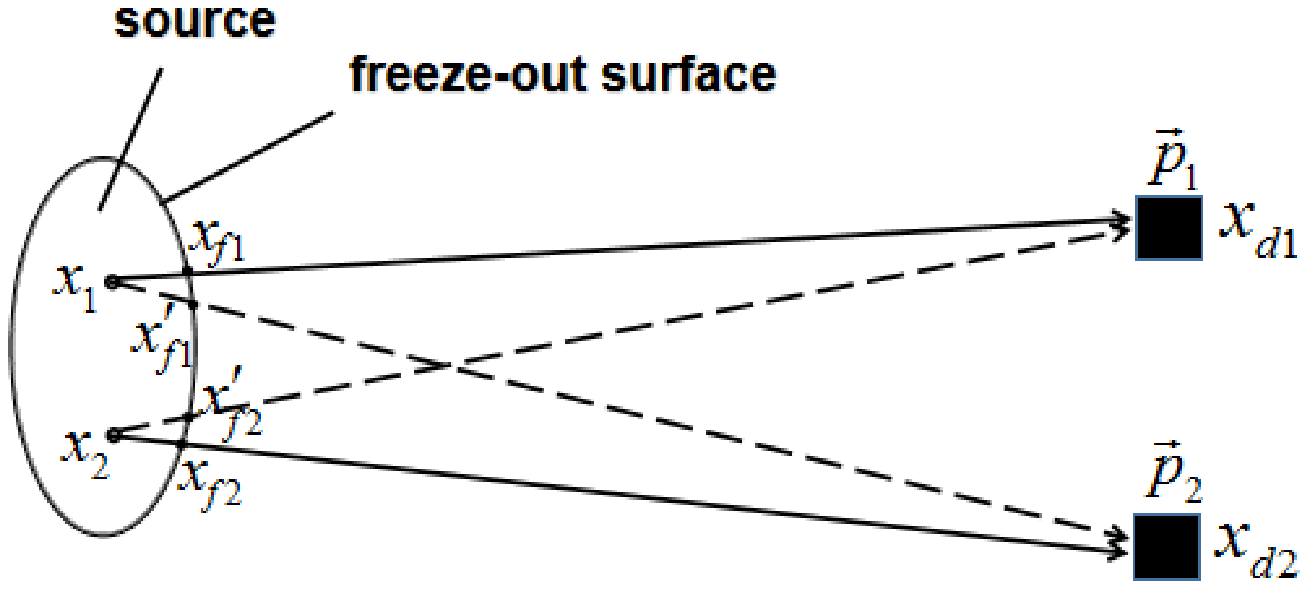}
\vspace*{-5mm}
\caption{Schematic diagram of two identical bosons produced at source points
$x_1$ and $x_2$, freeze-out at $x_{f1}$ and $x_{f2}$ (or $x'_{f1}$ and $x'_{f2}$),
and detected at $x_{d1}$ and $x_{d2}$ with momenta $\mk_1$ and $\mk_2$,
respectively. The solid lines joining $x_1 \to x_{f1} \to x_{d1}$ for $\mk_1$
and $x_2 \to x_{f2} \to x_{d2}$ for $\mk_2$, and the dashed-lines joining $x_1
\to x'_{f1} \to x_{d2}$ for $\mk_2$ and $x_2 \to x'_{f2} \to x_{d1}$ for $\mk_1$
are possible trajectories. }
\label{source}
\end{figure}

For the event that two identical bosons produced at source points $x_1$ and $x_2$
with four-momenta $p'_1$ and $p'_2$ and detected at $x_{d1}$ and $x_{d2}$ with
momenta $\mk_1$ and $\mk_2$, the possibility amplitude can be expressed with
quantum path-integral method as \cite{Wongbook},
\begin{eqnarray}
&&\psi_2(\mk_1\mk_2\!:x_1x_2\buildrel{m'_1m'_2}\over\longrightarrow x_{f1}x_{f2}
\buildrel{mm}\over\longrightarrow x_{d1}x_{d2})~~~~~~~~~~~~\nonumber\\
&&\hspace*{8mm}=\frac{1}{\sqrt{2}}\bigg[\psi_1(\mk_1\!: x_1 \buildrel
m'_1\over\longrightarrow x_{f1}\buildrel m\over\longrightarrow x_{d1})\nonumber\\
&&\hspace*{12mm}\times\,\psi_1(\mk_2\!: x_2 \buildrel m'_2\over\longrightarrow
x_{f2}\buildrel m\over\longrightarrow x_{d2})\nonumber\\
&&\hspace*{12mm}+\,\psi_1(\mk_1\!: x_2 \buildrel m'_1\over\longrightarrow x'_{f2}
\buildrel m\over\longrightarrow x_{d1})\nonumber\\
&&\hspace*{12mm}\times\,\psi_1(\mk_2\!: x_1 \buildrel m'_2\over\longrightarrow
x'_{f1}\buildrel m\over\longrightarrow x_{d2})\bigg].
\label{psi-2}
\end{eqnarray}
The two-boson possibility amplitude includes the two terms: one is for the
bosons detected at $x_{d1}$ and $x_{d2}$ from the source points $x_1$ and $x_2$
respectively as the solid-lines shown in Fig.~\ref{source}, and the other is for
the bosons detected at $x_{d1}$ and $x_{d2}$ from the source points $x_2$ and
$x_1$ respectively as the dashed-lines shown in Fig.~\ref{source}.

Considering the boson emitted in the source with a certain momentum and in-medium
mass, we can obtain for chaotic emission,
\begin{widetext}
\begin{eqnarray}
P_1(\mk)=\!\int\!dm' D(m')\bigg|\sum_x \psi_1(\mk\!:x\buildrel m'\over
\longrightarrow x_f \buildrel m\over\longrightarrow x_d)\bigg|^2
=\!\int\!dm' D(m')\sum_x {\cal A}^2(p',x)=\!\int\!dm' D(m') dx \rho(x)
{\cal A}^2(p',x),
\label{P-1}
\end{eqnarray}
where $D(m')$ is the in-medium mass distribution and $\rho(x)$ is the source
density distribution. And, we have
\begin{eqnarray}
&&\hspace*{-10mm}P_2(\mk_1,\mk_2)=\!\int\! dm'_1 dm'_2 D(m'_1) D(m'_2)\,
\frac{1}{2!}\bigg|\sum_{x_1,x_2}\psi_2(\mk_1
\mk_2\!:x_1x_2\buildrel{m'_1m'_2}\over\longrightarrow x_{f1}x_{f2}\buildrel{m
m}\over \longrightarrow x_{d1}x_{d2})\bigg|^2 \nonumber\\
%&&\hspace*{6mm}\,=\!\int\! dm'_1 dm'_2 D(m'_1) D(m'_2)\,
%\sum_{x_1,x_2}\frac{1}{2}\bigg|{\cal A}(p'_1,x_1)e^{ip'_1\cdot(x_1-x_{f1})}
%e^{ip_1\cdot(x_{f1}-x_{d1})} {\cal A}(p'_2,x_2)e^{ip'_2\cdot(x_2-x_{f2})}
%\nonumber\\
%&&\hspace*{10mm}\times\, e^{ip_2\cdot(x_{f2}-x_{d2})}
%+{\cal A}(p'_1,x_2)e^{ip'_1\cdot(x_2-x'_{f2})}e^{ip_1\cdot
%(x'_{f2}-x_{d1})}{\cal A}(p'_2,x_1)e^{ip'_2\cdot(x_1-x'_{f1})}e^{ip_2\cdot
%(x'_{f1}-x_{d2})}\bigg|^2 \nonumber\\
%&&\hspace*{6mm}\,=\!\int\! dm'_1 dm'_2 D(m'_1) D(m'_2)\,
%\sum_{x_1,x_2}\frac{1}{2}\Big\{{\cal A}^2(p'_1,x_1) {\cal
%A}^2(p'_2,x_2) + {\cal A}^2(p'_1,x_2) {\cal A}^2(p'_2,x_1) +{\cal A}(p'_1,x_1)
%\nonumber\\
%&&\hspace*{10mm}\times\, {\cal A}(p'_2,x_1) {\cal A}(p'_1,x_2){\cal A}(p'_2,x_2)
%\Big[e^{i(p'_1-p'_2)\cdot(x_1-x_2)}e^{i(p'_1-p_1)\cdot(x'_{f2}-x_{f1})}e^{i(p'_2
%-p_2)\cdot(x'_{f1}-x_{f2})}+c.c.\Big]\Big\} \nonumber\\
&&\hspace*{6mm}\,=P_1(\mk_1)P_1(\mk_2)+\!\int\! dm'_1 dm'_2 D(m'_1) D(m'_2)
dx_1 dx_2\,\rho(x_1)\rho(x_2) {\cal A}(p'_1,x_1) {\cal A}(p'_2,x_1) {\cal A}
(p'_1,x_2) {\cal A}(p'_2,x_2) \nonumber\\
&&\hspace*{36mm}\,\times {\rm Re} \Big[e^{i(p'_1-p'_2)\cdot(x_1-x_2)} e^{i(p'_1-p_1)\cdot(x'_{f2}-x_{f1})} e^{i(p'_2-p_2)\cdot(x'_{f1}-x_{f2})}\Big].
\label{P-2}
\end{eqnarray}
\end{widetext}
In above derivations, the terms that contain production phase $\phi$ cancel
out because of the randomness of the phases for chaotic sources \cite{Wongbook}.
Clearly, $P_2(\mk_1,\mk_2)$ will become the usual result for a chaotic source, 
if there is not the in-medium mass modification ($p'_1=p_1$, $p'_2=p_2$).

Assuming that the production amplitude is coordinate-independent in the source
and the bosons approximately freeze out at the same time, we have
\begin{eqnarray}
&&P_2(\mk_1,\mk_2)=P_1(\mk_1)P_1(\mk_2)+\!\int\! dm'_1 dm'_2 D(m'_1) D(m'_2)\cr
&&\hspace*{20mm}\times\,{\cal A}^2(p'_1) {\cal A}^2(p'_2) \big|\,\tilde
\rho(p'_1-p'_2)\big|^2.
\end{eqnarray}
For a Gaussian-distribution source $\rho(r)\!\sim\!e^{-t^2/2\tau^2-\mr^2/2R^2}$
and with the approximation, ${\cal A}(p')\approx {\cal A}(p)$, the two-particle
correlation function is
\begin{equation}
C(\mk_1,\mk_2)=1+I(\mk_1,\mk_2)\,e^{-|\mk_1-\mk_2|^2R^2},
\end{equation}
where
\begin{eqnarray}
I(\mk_1,\mk_2)=\!\int\! dm'_1 dm'_2 D(m'_1) D(m'_2)\,
e^{-(E'_{\mk_1}-E'_{\mk_2})^2 \tau^2}.
\end{eqnarray}

\begin{figure}[htbp]
\includegraphics[scale=0.55]{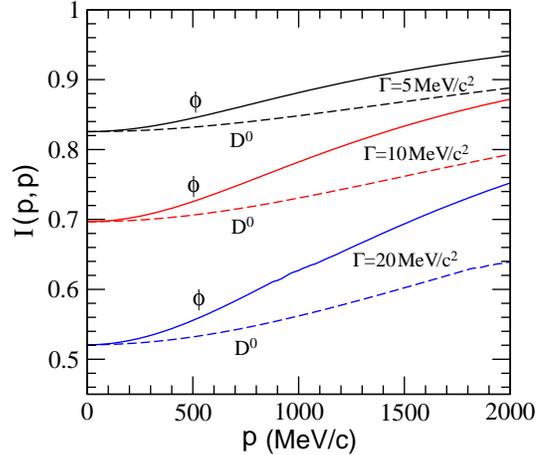}
\caption{(Color online) Strengths of HBT correlations of $\phi\phi$ (solid 
lines) and $D^0D^0$ (dashed lines) pairs. }
\label{zfIp}
\end{figure}

The strength of HBT correlation, $I(\mk,\mk)$, is related to the in-medium
mass distribution of the bosons. We plot in Fig.~\ref{zfIp} the strengths as
a function of particle momentum for $\phi$ and $D^0$ mesons. Here, we take
the source lifetime $\tau=10$~fm/$c$, and the mass distribution is taken to
be in Breit-Wigner form, $\frac{\Gamma/2\pi}{(m'-m-\Delta m')^2+\Gamma^2/4}$,
where $\Delta m'$ and $\Gamma$ are the in-medium mass shift and width
respectively. At $p=0$, the strength only depends on the width. Because the
average mass difference of two bosons in the medium is larger for larger
$\Gamma$ than that for smaller $\Gamma$, the strength decreases with increasing
$\Gamma$ value. When particle momentum increasing, $I(p,p)$ for $D^0$ meson
increases more slowly than that for $\phi$ meson, because $D^0$ has a heavier 
mass than $\phi$. All the results of $I(p,p)$ are independent of $\Delta m'$.

In high-energy heavy-ion collisions at the Relativistic Heavy Ion Collider (RHIC) 
and the Large Hadron Collider (LHC), the in-medium mass width of $\phi$ meson is 
about several MeV \cite{{NA49-PLB00p,STAR-PRC02p,NA50-PLB03p,PHENIX-PRC05p,
PHENIX-PRL07p,STAR-PRC09p,STAR-PLB09p,PHENIX-PRC11p,ALICE-PRC15p,PHENIX-PRC16p,
ALICE-EPJC18p}}. In this case, the influence of the in-medium mass width may 
decrease the strength of HBT correlation by about 10\% averagely. However, the 
in-medium mass width of $D^0$ meson in the collisions may reach several tens of 
MeV \cite{{ALICE-PRL13D,ALICE-PRC14D,ALICE-JHEP15D,ALICE-JHEP16D,ALICE-JHEP18D,
CMS-PLB18D,CMS-PRL18D,STAR-PRC19D}}. The suppression on the strength of the HBT 
correlation of $D^0$ meson is very significant.

In summary, the observed identical bosons may have different masses in the 
particle-emitting sources because of the interactions between the bosons and 
medium. In this letter, we have studied the influence of the in-medium mass 
width of identical bosons on the strength of HBT correlation. We find that 
the strength decreases with in-medium mass width of the identical bosons. 
This influence are more significant for the boson with heavier mass and 
decreases with increasing particle momentum. In high-energy heavy-ion collisions 
at the RHIC and LHC, the strength of HBT correlation of $\phi$ meson may reduce 
by about 10\% averagely. However, the influence on the HBT correlation of $D^0$ 
meson is very significant.

W. N. Zhang thanks Cheuk-Yin Wong for helpful discussions.
This research was supported by the National Natural Science Foundation of China
under Grant Nos. 11675034 and 11275037. \\

\end{document}